\documentclass[10pt,namedreferneces]{SolarPhysics}
\usepackage[optionalrh,solaenum]{spr-sola-addons} 

\usepackage{graphicx}
\usepackage{color}

\newcommand{\BE}{\begin{equation}}
\newcommand{\EE}{\end{equation}}
\newcommand{\BA}{\begin{eqnarray}}
\newcommand{\EA}{\end{eqnarray}}

\newcommand{\BIT}{\begin{itemize}}

\newcommand{\EIT}{\end{itemize}}

\newcommand{\brF}{ {\bar F}}
\newcommand{\tta}{ \theta_{1} }
\newcommand{\ttaa}{ \theta_{2} }

\def \half {\textstyle{\frac{1}{2}}}

\begin{document}
\begin{article}
\begin{opening}

\title{Evolution of Magnetic Helicity During Eruptive Flares and Coronal Mass Ejections}

\author{E. R.~\surname{Priest}$^1$\sep D.W.~\surname{Longcope}$^2$\sep M.~\surname{Janvier}$^3$$^4$}

\runningtitle{Magnetic helicity evolution during an eruptive flare}
\runningauthor{Priest, Longcope, \& Janvier}

\institute{$^1$ School of Mathematics and Statistics, University of St. Andrews, Fife KY16 9SS, Scotland, UK\\
$^2$ Dept. of Physics, Montana State University, Bozeman, MT, USA\\
$^3$ University of Dundee, School of Science and Engineering, DD1 4HN, UK\\
$^4$ Institut d'Astrophysique Spatiale, UMR8617, Univ. Paris-Sud-CNRS, Universit\'e Paris-Saclay, B\^atiment 121, 91405 Orsay Cedex, France \\
}

\begin{abstract}
During eruptive solar flares and coronal mass ejections, a non-pot{\-}ential magnetic arcade with much excess magnetic energy goes unstable and reconnects. It produces a twisted erupting flux rope and leaves behind a sheared arcade of hot coronal loops. We suggest that: the twist of the erupting flux rope can be determined from conservation of magnetic flux and magnetic helicity and equipartition of magnetic helicity. It depends on the geometry of the initial pre-eruptive structure.
Two cases are considered, in the first of which a flux rope is not present initially but is created during the eruption by the reconnection. In the second case, a flux rope is present under the arcade in the pre-eruptive state, and the effect of the eruption and reconnection is to add an amount of magnetic helicity that depends on the fluxes of the rope and arcade and the geometry. \end{abstract}


\date{Draft: \today}

\end{opening}

\section{Introduction} 
\label{sec1}
  
The standard understanding of eruptive solar flares (e.g., \opencite{schmieder12}; \opencite{priest14a}; \opencite{aulanier14}; \opencite{janvier15}) is that excess  magnetic energy and magnetic helicity build up until a threshold is reached at which point the magnetic configuration either goes unstable or loses equilibrium, either by breakout \cite{antiochos99a,devore08} or magnetic catastrophe \cite{demoulin88,priest90a,forbes91b,lin00,wang09} or kink instability \cite{hood79a} or by torus instability \cite{lin98,kliem06,demoulin10a,aulanier10,aulanier12}.

Some solar flares (known as {\it eruptive flares}) are associated with the eruption of a magnetic structure containing a prominence (observed as a coronal mass ejection) and typically produce a two-ribbon flare, with two separating H$\alpha$ ribbons joined by a rising arcade of flare loops.  Others are contained and exhibit no eruptive behaviour. While some coronal mass ejections are associated with eruptive solar flares,  others occur outside active regions and are associated with the eruption of a quiescent prominence.  Coronal mass ejections outside active regions do not produce high-energy products, because their magnetic and electric fields are much smaller than in eruptive solar flares, but their magnetic origin and evolution may well be qualitatively the same.

Magnetic helicity is a measure of the twist and linkage of magnetic fields, and its basic properties were developed by \opencite{woltjer58}, \opencite{taylor74}, \opencite{moffatt78},  \opencite{berger84b}, \opencite{berger00} and \opencite{demoulin06b}. 
It was first suggested to be important in coronal heating, solar flares and coronal mass ejections by \opencite{heyvaerts84}, who proposed that, when the stored magnetic helicity is too great, it may be ejected from the Sun in an erupting flux rope (see also \opencite{rust95}; \opencite{low03}; \opencite{kusano02b}). 
The  flux of magnetic helicity through the photosphere, its buildup in active regions and its relation to sigmoids has been studied by \opencite{pevtsov95}, \opencite{canfield98}, \opencite{canfield99,canfield00}, \opencite{pevtsov00b}, \opencite{pevtsov02a}, \opencite{green02a}, \opencite{pariat06b}, and \opencite{poisson15}. 

Indeed, the measurement of magnetic helicity in the corona is now a key topic with regard to general coronal evolution \cite{chae01a,demoulin02a,mandrini04a,zhang06,zhang08,mackay11,mackay14,gibb14}. Also, since magnetic helicity is well-conserved on timescales smaller than the global diffusion time, measuring it in interplanetary structures such as flux ropes and magnetic clouds \cite{gulisano05,qiu07,hu14,hu15} allows us to link the evolution of CMEs in the solar wind with their source at the Sun \cite{nindos03,luoni05}.

The common scenario described above for an eruptive solar flare or coronal mass ejection is that, after the slow buildup of magnetic helicity in a magnetic structure, the eruption is triggered and drives three-dimensional reconnection which adds energy to post-flare loops.  Originally, it was thought that all of the magnetic energy stored in excess of potential would be released during a flare, and therefore that the final post-flare state would be a potential magnetic field.  However, the modern realisation is that magnetic helicity conservation provides an extra constraint that produces a different final state. It is also observed that flare loops in an eruptive flare do not relax to a potential state, since the low-lying loops remain quite sheared \cite{asai04,warren11c,aulanier12}. Our aim in this paper is to determine two key observational consequences of the reconnection process, namely, the amount of magnetic helicity, and therefore twist, in the erupting flux rope, and in the shear of the underlying flare loops. 

\begin{figure}[h]
{\centering
 \includegraphics[width=9cm]{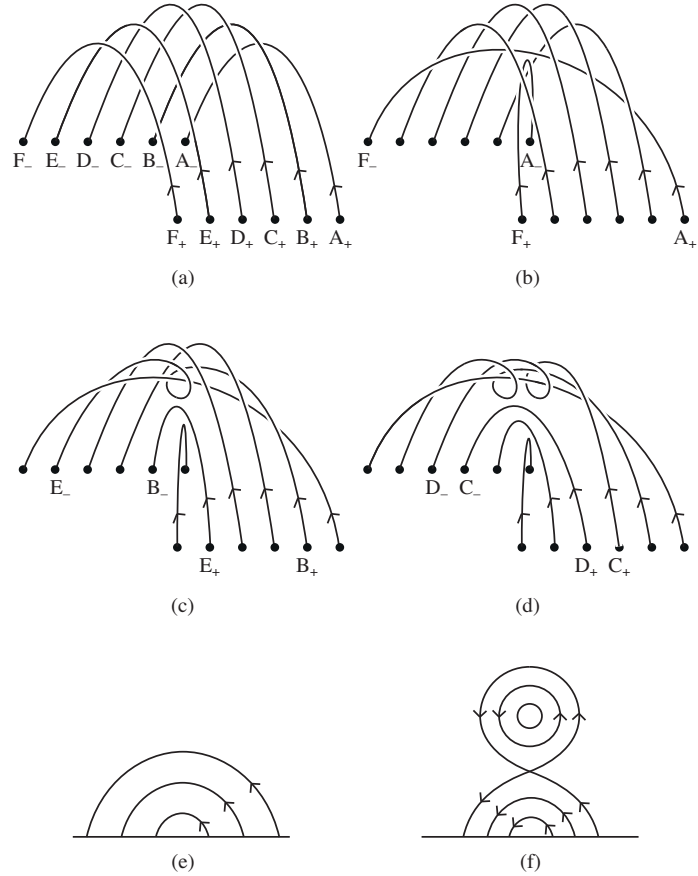}
\caption{(a) A  simple sheared arcade that reconnects to produce (d) an erupting flux rope plus an underlying less-sheared arcade. In the initial state (a), footpoints A$_+$, B$_+$, C$_+$, D$_+$, E$_+$, F$_+$, are joined  to footpoints A$_-$, B$_-$, C$_-$, D$_-$, E$_-$, F$_-$, respectively.  In the first stage, shown in (b), loops A$_+$A$_-$ and F$_+$F$_-$ have reconnected to give new loops A$_+$F$_-$ and F$_+$A$_-$.  In the second stage, shown in (c), loops B$_+$B$_-$ and E$_+$E$_-$ have reconnected to give new loops  B$_+$E$_-$ and E$_+$B$_-$.  Then in the final stage, shown in (d), loops C$_+$C$_-$ and D$_+$D$_-$ have reconnected to give a twisted flux rope  C$_+$D$_-$ that may erupt and a set of underlying loops D$_+$C$_-$.  (e) and (f) show a projection of the field lines onto a vertical section through the configuration, looking from the left, in the initial (left) and final state (right) during the eruption.}
\label{fig1}}
\end{figure}
\begin{figure}[h]
{\centering
 \includegraphics[width=12cm]{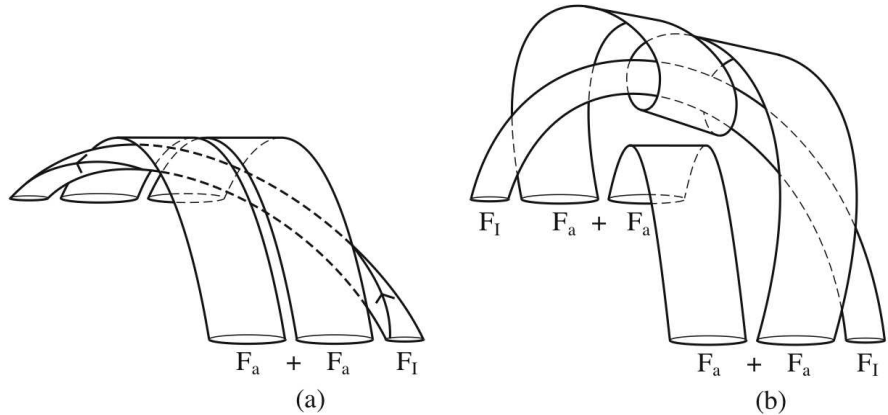}
\caption{(a) A sheared arcade overlying a flux rope reconnects to produce (b) an erupting flux rope plus an underlying less-sheared arcade.}
\label{fig2}}
\end{figure}
We determine the magnetic helicity in the erupting flux rope and underlying flare loops in two cases. In the first  (Fig.\ref{fig1}), no flux rope is present in the initial arcade, but a flux rope is created during the process of the reconnection.  In the second case (Fig.\ref{fig2}), the initial state consists of a magnetic arcade that overlies a flux rope, and the effect of the eruption and reconnection is to enhance the flux and twist of the flux rope. Our aim is to produce the simplest model that preserves the core physics of the process.  For example, we neglect the internal structure of the coronal arcade and model it simply as a  flux rope whose photospheric footpoints are stretched out in two lines either side of the polarity inversion line. The second case (with the initial flux rope) is much more likely to go unstable and erupt.

We calculate the twist in the erupting flux rope  from the properties of the initial pre-eruptive state by making three assumptions, namely:

(i) conservation of magnetic flux;

(ii) conservation  of magnetic helicity;

(iii) and equipartition magnetic helicity.

The third assumption implies that the same amount of magnetic helicity is transferred by  reconnection to the  erupting flux rope and underlying arcade \cite{wright89}.  We have considered an alternative possibility, namely, preferential transfer of magnetic helicity to the flux rope during the reconnection, which would imply the flaring loops have vanishing self-helicity. However, this seems less likely,  
because, although the flare loops are generally seen as non-twisted structures, they are also observed to be nonpotential, since the low-lying loops appearing early in the flare possess more shear than the high-lying loops \cite{aulanier12}.

The reconnection of twisted tubes has been studied in landmark papers by \opencite{linton01}, \opencite{linton02}, \opencite{linton03}, \opencite{linton05} in which they  consider also the extra constraint of energy both numerically and analytically. Straight tubes of a variety of twists and inclinations are brought together by an initial stagnation-point flow and allowed to reconnect self-consistently by either bouncing, slingshot, merging or tunnelling reconnection. Although energy is not considered in detail here, we make initial comments about energy considerations in Section \ref{sec3.3} and hope to develop them further in future numerical treatments.

The structure of the paper is as follows. In Section \ref{sec2} we summarise the basic properties of magnetic helicity that are needed for our analysis. Then in Section \ref{sec3} we present our simple model for  a sheared magnetic arcade in which no flux rope is present initially but is created during the eruption by the reconnection.   This is followed in Section \ref{sec4} by a different model in which the initial arcade contains a flux rope, which erupts and  leaves behind a less-sheared arcade.  Finally, suggestions for follow-up are given in  Section \ref{sec5}.

\section{Magnetic Helicity Preliminaries}
\label{sec2}

Magnetic helicity is a topological quantity that comprises two parts: the {\it self-helicity} measures the twisting and kinking of a flux tube, whereas the {\it mutual helicity} refers to the linkage between different flux tubes \cite{berger86,berger99}. Their sum, the relative helicity, is a global invariant that is conserved during ideal evolution and that decays extremely slowly (over the global magnetic diffusion time, $\tau_d$) in a weakly resistive medium -- i.e., one for which the global magnetic Reynolds number is large ($R_m\gg 1$). Thus, during magnetic reconnection in the solar atmosphere over a timescale $t\ll \tau_d$, the total magnetic helicity is approximately conserved, but it may be converted from one form to another, say, from mutual to self.

If a magnetic configuration consists of $N$ flux tubes, its total magnetic helicity may be written in terms of the self-helicity ($H_s$) of each tube due to its own internal twist and the mutual helicity ($H_m$) due to the linking of the tubes as
\begin{eqnarray}
H=\Sigma_{i=1}^N H_{si}+2\Sigma_{i<j=1}^N H_{mij}.
\nonumber
\end{eqnarray}

\begin{figure}[h]
{\centering
 \includegraphics[width=12cm]{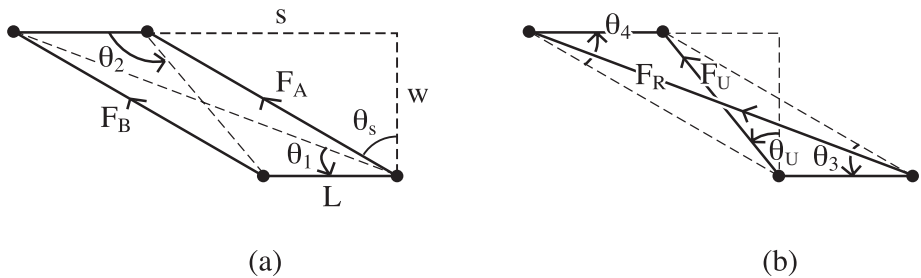}
\caption{The basic configurations seen from above:  (a)  two flux tubes of flux $F_A$ and $F_B$ side by side and (b) one flux tube of flux $F_R$ crossing over another underlying tube of flux $F_U$.}
\label{fig3}}
\end{figure}
The basic configurations used in this paper are a single flux rope of magnetic flux $F$, say, and a pair of flux tubes side by side with fluxes $F_A$ and $F_B$.  The twist $\Phi(F)$ of the flux rope is in general not uniform, but a mean twist $\bar{\Phi}(F)$ may be defined in an appropriate way (see Sec.\ref{sec3.2.2}). Then the self-helicity of the flux rope is
\begin{equation}
H_s=\frac{{\bar \Phi}}{2\pi}F^2,
\label{eq1}
\end{equation}
while the mutual helicity of the two tubes side by side is 
\begin{equation}
H_m=\frac{\theta_{2}-\theta_{1}}{2\pi}F_{A}F_{B}, 
\label{eq2}
\end{equation}
where the angles $\theta_1$ and $\theta_2$ depend on the geometry shown in Fig.\ref{fig3}a \cite{berger98,demoulin06b}. In contrast, a pair of crossing flux tubes, one of which has flux $F_R$ overlying the other underlying tube of flux $F_U$, say,  possesses  mutual helicity
\begin{equation}
H_m=-\frac{\theta_{3}+\theta_{4}}{2\pi}F_{R}F_{U},
\label{eq3}
\end{equation}
where the angles $\theta_3$ and $\theta_4$ depend on the geometry of the footpoints in Fig.\ref{fig3}b. In the models that follow, the overlying tube will represent an erupting flux rope, whereas the underlying tube will model an arcade of flare loops, which is why we have used the notations $F_R$ and $F_U$. Also, for simplicity, the footpoints will form a parallelogram shape and so $\theta_4=\theta_3$. Note that the sign in Eq.(\ref{eq3})   is determined by a right-hand rule in the sense that, if the fingers of the right hand are directed along the overlying tube, then the sign is positive if the underlying tube is in the direction of the thumb or, as in the case of Fig.\ref{fig3}b, the sign is negative if the underlying tube is in the opposite direction. Thus, if the direction of flux in either tube is reversed,  the helicity is multiplied by $-1$.  If the tube of flux $F_R$ passes under the other tube rather than over it, then $\theta_3+\theta_4$ is replaced by $\theta_3+\theta_4-2\pi$. 

\section{Modelling the Eruption of a Simple Magnetic Arcade with No Flux Rope} 
\label{sec3}

Consider  a simple sheared arcade (Fig.\ref{fig1}a),  which reconnects and produces an erupting flux rope together with an underlying arcade  (Fig.\ref{fig1}b). Our aim is to deduce the helicity  in the flux rope (and so its twist) in terms of the shear and helicity of the initial arcade.  In addition, we shall deduce the distribution of twist within the flux rope.

\subsection{Overall Process}
\label{sec3.1}

We first consider the overall process and model the configuration in the simplest way by treating the initial arcade as two untwisted flux tubes side by side. We compare the initial state in Fig.\ref{fig3}a  having two untwisted flux tubes (of equal flux $F_A=F_B=F_a$) side by side with the reconnected state in Fig.\ref{fig3}b having a twisted erupting flux rope (of mean twist ${\bar \Phi}$, say) overlying flare loops.  

The initial configuration possesses mutual helicity ($H^i$), which may be calculated from Eq.(\ref{eq2})  as
\begin{equation}
H^i=\frac{\theta_{2}-\theta_{1}}{\pi}F_a^2,
\label{eq4}
\end{equation}
where the angles $\theta_{1}$ and $\theta_{2}$   can be written in terms of the width ($w$), length ($L$) and shear ($s$) of the arcade as
\begin{equation}
\tan \theta_1=\frac{w}{L+s}=\frac{1}{\bar{L}+\bar{s}}, \ \ \ \ \ \tan \theta_2=\frac{w}{L-s}=\frac{1}{\bar{L}-\bar{s}},
\label{eq5}
\end{equation}
with dimensionless length and shear being defined as
\begin{eqnarray}
\bar{L}=\frac{L}{w} \ \ {\rm and} \ \  \bar{s}=\frac{s}{w}.
\nonumber
\end{eqnarray}
As the shear $\bar{s}$ increases from zero to infinity, $\theta_1$ decreases from $ \pi/4$ (when $\bar L=1$) to zero, while $\theta_2$ increases from $\pi/4$ to $\pi$.

During the reconnection process we assume {\it flux conservation}, so that
 the total final flux ($F_R+F_U$) equals the total initial flux ($2Fa$). We also assume that reconnection feeds flux simultaneously and equally into the rope and underlying loops, so that $F_R=F_U$. The net result is that in the reconnected configuration (Fig.\ref{fig3}b) the fluxes of the erupting flux rope and underlying loops are both equal to half the initial total flux ($F_R=F_U=F_a$). We also assume {\it helicity equipartition}, so that the released mutual helicity is added equally to the erupting flux rope and underlying loops. Then the magnetic helicity in the reconnected state, namely, the sum of the self-helicities and mutual helicity of the flux rope and  the underlying sheared loops, may be written for simplicity (assuming $\theta_4=\theta_3$) as
\begin{equation}
H^r=\frac{{\bar \Phi}}{\pi}F_a^2-\frac{2\theta_{3}}{\pi}F_a^2,
\label{eq6}
\end{equation}
where
\begin{eqnarray}
\tan \theta_3=\frac{1}{\bar{s}},
\nonumber
\end{eqnarray}
while the self-helicity of the underlying sheared loops is written for simplicity in the same form as that of a twisted structure, namely,
\begin{eqnarray}
H_{sU}=\frac{{\bar \Phi}}{2\pi}F_a^2.
\nonumber
\end{eqnarray}
As the shear $\bar{s}$ increases from zero to infinity, $\theta_3$ decreases from $\half \pi$ to zero.

\begin{figure}[h]
{\centering
 \includegraphics[width=12cm]{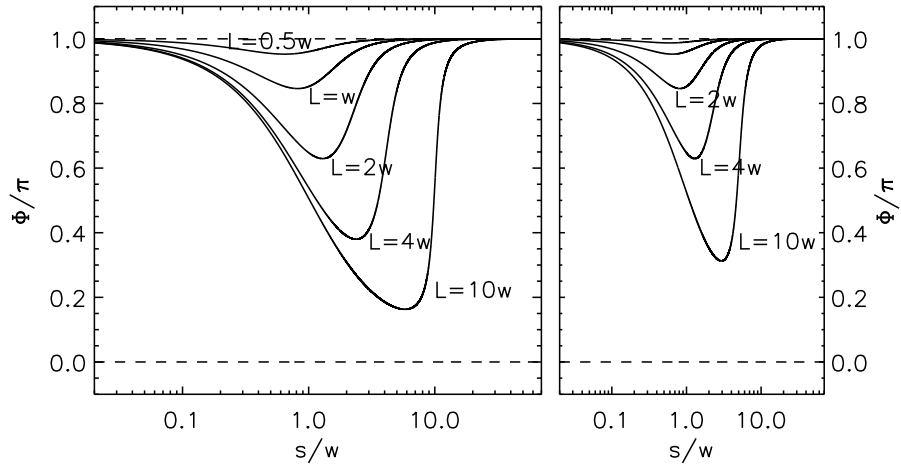}
\caption{The mean twist (${\bar \Phi}$) of the erupting flux rope  as a function of dimensionless shear $\bar{s}=s/w$ for several values of the dimensionless arcade length $\bar{L}=L/w$ in terms of the arcade width ($w$) for (a) the simple overall model and (b) for the evolutionary model (Sec.\ref{sec3.2}) when a fraction $F/F_a=\half$ of the initial flux $F_a$ has reconnected.}
\label{fig4}}
\end{figure}
Then magnetic helicity conservation (equating $H^i$ from Eq.(\ref{eq4}) with $H^r$ from Eq.(\ref{eq6})) determines the mean flux-rope twist as
\begin{equation}
{\bar \Phi}=\theta_2-\theta_1+2\theta_3,
\label{eq7}
\end{equation}
or, after substituting for the angles and manipulating,
\begin{equation}
{\tan  {\bar \Phi}}=-\frac{2\bar{s}\bar{L}^2}{(1+\bar{s}^2)^2-(\bar{s}^2-1)\bar{L}^2}.
\label{eq8}
\end{equation}
If instead the released mutual helicity is added preferentially to the flux rope, then this value of ${\bar \Phi}$ is doubled.

The way in which the twist varies with shear ($\bar{s}$) and arcade length ($\bar{L}$) is shown in Fig.\ref{fig4}a. Thus, for example, when $s\ll L= w$, then $\theta_1\approx\theta_2=\pi/4$ and $\theta_3= \pi/2$ and so ${\bar \Phi} = \pi$, whereas, when $s=w\ll L$, then $\theta_1\approx \theta_2\approx \pi/8$ and $\theta_3=\pi/4$ and so ${\bar \Phi} \approx \half\pi$. On the other hand, when $s\gg w=L$, then $\theta_1\approx \theta_3\approx 0$ and $\theta_2\approx \pi$ and so ${\bar \Phi} \approx \pi$.

Note that the initial mutual helicity (Eq.\ref{eq4}) is positive, since $\theta_2>\theta_1$, whereas the final mutual helicity (the second term in Eq.\ref{eq6}) is negative. Thus, equating the initial and final helicities to give Eq.\ref{eq7} implies that the self-helicity of the flux rope must be positive and so the twist has to be positive as indicated in Figs.\ref{fig1} and \ref{fig5}c (see also Figs.\ref{fig2} and \ref{fig9}a).  In other words, the effect of the reconnection is to decrease the mutual helicity and increase the self-helicity.  

The  shear angles  of the initial arcade ($\theta_s$) and of the underlying flare loops ($\theta_U$) are given by 
\begin{eqnarray}
\tan \theta_s = \frac{s}{w}\ \ \ \ \ \ \ \tan \theta_U =\frac{s-L}{w},
\nonumber
\end{eqnarray}
so that the flare loop shear is smaller than the initial loop shear (Fig.\ref{fig3}). If this were applied to a sequence of reconnecting arcade loops that are progressively further and further apart, it can be seen that the flare loop shear angle $\theta_U$ decreases as $w$ increases.

\subsection{Evolution of the Process}
\label{sec3.2}

\subsubsection{Using Helicity Conservation to Deduce the Mean Flux Rope Twist}
\label{sec3.2.1}

\begin{figure}[h] 
{\centering
 \includegraphics[width=12cm]{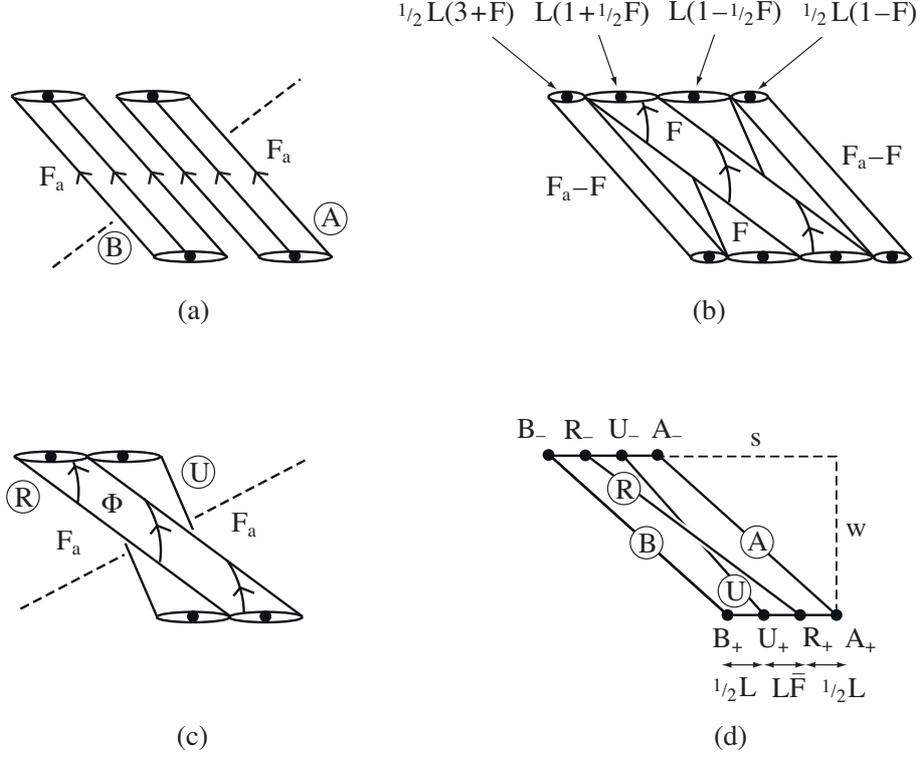}
\caption{The notation for our model of two untwisted flux tubes (A and B, both of flux $F_a$) side by side that produce an erupting flux rope (R) of flux $F$ and twist ${\bar \Phi}$ together with an underlying arcade of loops (U) of flux $F$. (a) shows the initial state, (b) an intermediate reconnected state, (c) the final reconnected state and (d) the notation of the geometry.}
\label{fig5}}
\end{figure}

Next, we consider the evolution of the process by again supposing the initial configuration consists of two untwisted flux tubes stretching from fixed photospheric flux sources to sinks (Fig.\ref{fig5}a).  The reconnected state again consists of an erupting twisted flux rope (R) and underlying arcade of loops (U) (Fig.\ref{fig5}c).

During reconnection, in going from Fig.\ref{fig5}a to Fig.\ref{fig5}c via intermediate states of the form Fig.\ref{fig5}b, we assume {\it conservation of magnetic flux}, so that, if the new flux rope (R) joining A$_+$ to B$_-$ gains flux $F$, then so does the underlying arcade (U) joining B$_+$ to A$_-$, while both flux tubes (A) and (B) of flux $F_a$  lose flux $F$ so that their fluxes both become $F_a-F$.
We suppose the footpoints form a parallelogram, so that  $\theta_3=\theta_4$.  
Initially, the arcade is modelled as consisting of two flux tubes (A and B) side by side, the centres of whose footpoints are indicated by large dots separated by a distance $L$ (Fig.\ref{fig5}a). During the course of the reconnection (Fig.\ref{fig5}b), when a fraction $\bar{F}=F/F_a$ of the flux has been reconnected, there is an overlying flux rope (R) of flux $F$, the centres of whose footpoints are located at R$_+$ and R$_-$, together with a underlying flux loop (U) with footpoints at U$_+$ and U$_-$ (Fig.\ref{fig5}d).  Also, the remaining unreconnected flux consists of two flux tubes (A and B) whose footpoints are located at A$_+$,  A$_-$ and B$_+$,  B$_-$, respectively (Fig.\ref{fig5}d). The centres of the upper ends of the four flux tubes are located at distances $\half L(1-\bar{F})$, $L(1-\half \bar{F})$, $ L(1+\half\bar{F})$ and $\half L(3+\bar{F})$, respectively, from the right-hand end of the arcade (Fig.\ref{fig5}c), and so they are separated by distances $\half L$, $L\bar{F}$ and $\half L$, as shown in Fig.\ref{fig5}d.  The edges of the upper ends of the tubes are located at distances $L(1-\bar{F})$, $L$, $ L(1+\bar{F})$ and $2L$ from the right-hand end of the arcade.

The second assumption is {\it magnetic helicity conservation}, so that the total magnetic helicity (i.e., the sum of self and mutual helicity) is conserved during the reconnection process, namely,
\begin{eqnarray}
H_{ARUB}^i=H_{ARUB}^r,
\nonumber
\end{eqnarray}
where $H_{ARUB}^i$ and $H_{ARUB}^r$ are the initial and reconnected helicities when a fraction $\bar{F}$ of flux has been reconnected, which are calculated below.
Initially, we assume two untwisted flux tubes side by side (or equivalently four untwisted flux tubes A, R, U and B), which possess mutual helicity (given by sums of terms of the form Eq.(\ref{eq2})) but no self-helicity.  During and after reconnection, the flux rope (R) and underlying loops (U) possess self-helicity ($H_{SR}$ and $H_{SU}$), but also there are mutual helicities ($H_{AB},H_{AR},H_{AU}$)  between loop A and loops  R and U,  ($H_{BR},H_{BU}$) between loop B and loops R and U, and ($H_{RU}$) between R and U, so that the  magnetic helicity after reconnection has the form
\begin{equation}
H_{ARUB}^r=H_{SR} +H_{SU}+2H_{AB} + 2H_{AR} + 2H_{AU} + 2H_{BR} + 2H_{BU} + 2H_{RU},
\label{eq9}
\end{equation}

The initial helicity is given by assuming the initial situation consists of four flux tubes of fluxes $F_a-F,\ F,\ F$ and  $F_a-F$ aligned  parallel to one another and stretching between $A_+$ and $A_-$, $R_+$ and $U_-$, $U_+$ and $R_-$, $B_+$ and $B_-$ in Fig.\ref{fig5}d. Thus, it consists of the sums of the mutual helicities of these tubes, namely,
\begin{eqnarray}
H_{ARUB}^i=2H^{i}_{AB}+2H^i_{AR}+2H^i_{AU}+2H^i_{BR}+2H^i_{BU}+2H^i_{RU},
\label{eq10}
\end{eqnarray}

The third assumption is {\it magnetic\ helicity\ equipartition}, so that the effect of reconnection of A and B is to give self-helicity equally to  the flux rope (R) and underlying loops (U), i.e.,
\begin{eqnarray}
H_{SU}=H_{SR}=\frac{{\bar \Phi}}{2\pi}F^2.
\label{eq11}
\end{eqnarray}

The expressions for the mutual helicities are given in the Appendix.  By substituting them into the expressions for initial and post-reconnection helicity,  helicity conservation  determines the flux rope self-helicity (and therefore its mean twist, ${\bar \Phi}$).  It transpires that,  although all of the mutual helicities change during the course of the reconnection, the sums $H_{AR}+H_{AU}$ and $H_{BR}+H_{BU}$ remain the same.  In other words, the unreconnected parts of flux tubes A and B are spectators, since they have not taken part in the reconnection, and we just need to take account of the other parts and the way they reconnect to give flux tubes R and U in Fig.\ref{fig5}d.  But this process is exactly what we have already considered in Sec.\ref{sec3.1} with one difference, namely, that the length $\bar L$ is replaced by $\bar{L}\bar{F}$ and the angles $\theta_1$ and $\theta_2$ are replaced by $\theta_1^{RU}$ and $\theta_2^{RU}$, respectively.

By using Eqns.(\ref{eq5}), (\ref{eq7}) and (\ref{eq8}), the resulting mean twist therefore becomes (see the Appendix)
\begin{eqnarray}
{\bar \Phi} = \theta_2^{RU}-\theta_1^{RU}+2\theta_3,
\label{eq12}
\end{eqnarray}
where 
\begin{eqnarray}
\tan\theta_1^{RU}=\frac{1}{\bar{L}\bar{F}-\bar{s}},\ \ \ \ \ \ \tan\theta_2^{RU}=\frac{1}{\bar{L}\bar{F}+\bar{s}} \ \ \ \ \ \ \ \tan \theta_3=\frac{1}{\bar{s}}.
\label{eq13}
\end{eqnarray}
After manipulating this gives
\begin{eqnarray}
{\tan{\bar \Phi}}=-\frac{2\bar{s}\bar{L}^2\bar{F}^2}{(1+\bar{s}^2)^2-(\bar{s}^2-1)\bar{L}^2\bar{F}^2},
\label{eq14}
\end{eqnarray}

The mean twist is plotted in Fig.\ref{fig4}b.  When $\bar{F}\ll1$ this reduces to
\begin{eqnarray}
{\bar \Phi}=-\frac{2\bar{s}\bar{L}^2\bar{F}^2}{(1+\bar{s}^2)^2},
\nonumber
\end{eqnarray}
whose magnitude increases with $\bar{s}$ from zero to a maximum of $12\sqrt 3\bar{L}^2\bar{F}^2/16$ at $\bar{s}=1/\sqrt 3$ (corresponding to a shear angle of $\pi/3$) and then declines to zero.

For complete reconnection, the flux tube has eaten its way right through the overlying arcade and so ${\bar F}=1\ (F=F_a)$. The resulting final mean flux rope twist (${\bar \Phi}_f$) is given by
\begin{equation}
{\tan{\bar \Phi}_f}=-\frac{2\bar{s}\bar{L}^2}{(1+\bar{s}^2)^2-(\bar{s}^2-1)\bar{L}^2},
\label{eq15}
\end{equation}
which is the same as Eq.(\ref{eq8}) and is plotted as a function of shear for various arcade aspect ratios in Fig.\ref{fig4}a.

\begin{figure}[h]
{\centering
 \includegraphics[width=12cm]{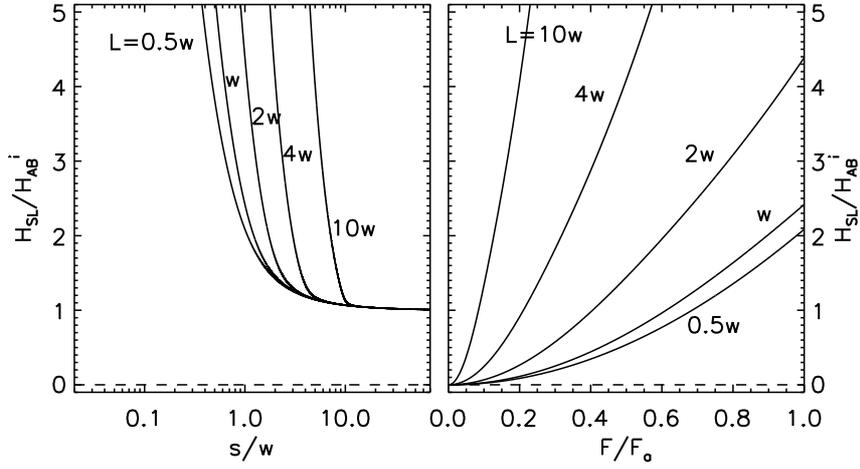}
\caption{The magnetic helicity of the underlying arcade of loops (a) as a function of shear (s) when the reconnection is complete and (b) as a function of reconnected flux for several arcade geometries when $s=w$.}
\label{fig6}}
\end{figure}

Another quantity of interest is the magnetic helicity of the underlying flare loops [$H_{sU}={\bar \Phi} F^2/(2\pi)$] as a fraction of the initial helicity [$H_{AB}^i=(\ttaa^i-\tta^i)F_a^2/(2\pi)$] given by Eq.(\ref{eq4}). After using Eq.(\ref{eq12}) for ${\bar \Phi}$, this may be written
\begin{eqnarray}
\frac{H_{sU}}{H_{AB}^i}=\frac{(\theta_2^{RU}-\theta_1^{RU}+2\theta_3){\bar F}^2}{(\theta_2-\theta_1)},
\nonumber
\end{eqnarray}
which is plotted in Fig.\ref{fig6}a as a function of reconnected flux ($F$) for a unit shear ($s=w$) and several arcade aspect ratios ($L/w$),  showing how the arcade helicity increases with reconnected flux.
An alternative to assuming helicity equipartition would be to suppose the self-helicity is transferred preferentially to the flux rope, so that the the self-helicity  ($H_{SU}$) of the underlying loops vanishes. The effect of this would be to double the value of the flux rope twist.

\subsubsection{Deducing the Distribution of Twist within the Flux Rope}
\label{sec3.2.2}

Using the dependence of mean twist $(\bar \Phi)$ of a flux rope on flux ($F$) it is possible to deduce the dependence of the twist itself ($\Phi)$ on flux. The mean twist is defined in such a way that the self-helicity of the flux rope satisfies Eq.(\ref{eq1}), namely,
\begin{eqnarray}
H_s=\frac{\bar \Phi}{2\pi} F^2.
\nonumber
\end{eqnarray}
Suppose the rope consists of a set of nested flux surfaces and $F$ is the axial flux within a particular flux surface. If the twist is not uniform but varies with flux or with distance from the rope axis, the self-helicity of the flux rope may be written \cite{berger98}
\begin{eqnarray}
H_s=\frac{1}{\pi} \int_0^F \Phi (F) F dF,
\nonumber
\end{eqnarray}
which reduces to $H_s=\Phi F^2/(2\pi)$ for a uniform twist.
Equating these two expressions for $H_s$, we find
\begin{eqnarray}
{\bar \Phi}=\frac{2}{F^2}\int \Phi(F)FdF,
\label{eq16}
\end{eqnarray}
which gives the definition of  average twist. It may be inverted to give the twist as a function of flux in terms of the mean twist, namely,
\begin{eqnarray}
\Phi(F)=\frac{1}{2F}\frac{d}{dF}\left(F^2 {\bar \Phi}\right).
\nonumber
\end{eqnarray}

\begin{figure}[h]
{\centering
 \includegraphics[width=12cm]{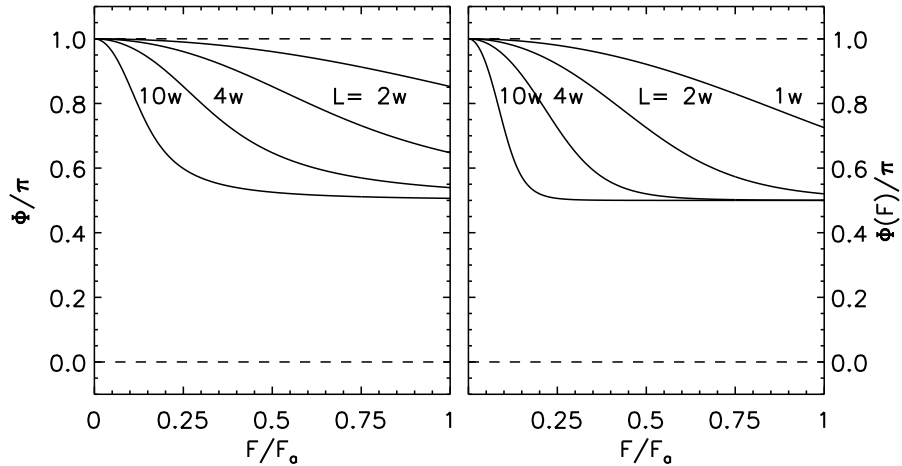}
\caption{(a) The mean flux rope twist (${\bar \Phi}$) as a function of reconnected flux ($\brF=F/F_a$) and (b)  flux rope twist as a function of flux ($\brF$) for a shear value $s/w=1$ and several values of the ratio ($L/w$) of arcade length to width.}
\label{fig7}}
\end{figure}
In terms of $\bar F$, this becomes
\begin{eqnarray}
\Phi({\bar F})=\frac{1}{2{\bar F}}\frac{d}{d{\bar F}}\left({\bar F}^2 {\bar \Phi}\right)={\bar \Phi}+{\bar F}^2\frac{d{\bar \Phi}}{d{\bar F}^2}.
\label{eq17}
\end{eqnarray}
By assuming that the average twist and enclosed axial flux obey Eq.(\ref{eq14}) at every flux surface within the flux rope, we may substitute
 for $\Phi({\bar F})$ from Eq.(\ref{eq14})  and deduce the distribution of twist within the erupting flux rope  as
\begin{eqnarray}
\Phi({\bar F}) &=&{\tan^{-1}}\left[-\frac{2{\bar s}{\bar L}^2{\bar F}^2}{(1+{\bar s}^2)^2-({\bar s}^2-1){\bar L}^2{\bar F}^2}\right]\nonumber\\[3pt]
&~~&~~-\frac{2{\bar s}{\bar L}^2{\bar F}^2}{(1+{\bar s}^2)^2-2({\bar s}^2-1){\bar L}^2{\bar F}^2+{\bar L}^4{\bar F}^4},
\label{eq18}
\end{eqnarray}
which is plotted in Fig.\ref{fig7}b.  From Fig.\ref{fig7}a we can see that the mean twist ($\bar \Phi$) decreases with reconnected flux $\bar F$ from $\pi$ to a value at ${\bar F}=1$ that lies between $\pi$ and $\half \pi$, and that this fully reconnected value decreases with increasing arcade length $\bar L$.  Eq.(\ref{eq17}) and Fig.\ref{fig7}b show that the twist itself $(\Phi)$  also starts from $\pi$ at ${\bar F}=0$ and lies between $\pi$ and $\half \pi$. Furthermore, since $d {\bar \Phi}/d{\bar F}^2<0$, $\Phi$ is smaller than $\bar \Phi$ and is roughly constant over most of the flux rope cross-section, especially when ${\bar L}>4$.

\subsection{Energy Considerations}
\label{sec3.3}

If the initial state is driven to reconnect by footpoint motions, then there is no need for the initial magnetic energy to exceed the final energy, since any change in energy could come from the work done by the footpoints.  If, however, the reconnection arises from some kind of instability, then an extra constraint that we have not considered so far arises from the condition that the magnetic energy of the initial state must exceed that of the final state. This will then rule in some of the changes we have considered and rule out others.  We do not here give a definitive answer to which changes are allowed, but only undertake a preliminary investigation. In Sec. \ref{sec3.3.1} we first consider a general expression for the free energy and demonstrate that there is indeed energy release during reconnection to start with, which implies that the magnetic helicity transfers we have been considering so far provide an upper limit on what will happen in reality.  Then in  Sec. \ref{sec3.3.2} we describe a simple model for calculating the free energy and again demonstrate that in some cases the free energy is indeed positive. In future, we look to a full numerical solution to provide a more in-depth account of the constraints produced by energy considerations.

Two minor points are worth making first. It may naturally be thought that two parallel untwisted tubes side by side (with flux going from footpoints A$_+$ and B$_+$ to A$_-$ and B$_-$, respectively) will not reconnect. This is indeed true if the initial state is potential and the footpoints form a rectangle, with reflectional symmetry, since the initial field is a state of minimum possible energy (e.g., Longcope, 1996).  In our case, however, the footpoints form a non-rectangular parallelogram lacking this symmetry and the initial state is force-free (Figure \ref{fig3}a), so there exists a lower-energy (potential) state with the same footpoints but with flux going from A$_+$ to B$_-$ as well as to B$_+$; some reconnection is therefore likely to occur.  However, the minimum-energy state that has given footpoint fluxes and given total magnetic helicity is not the potential state but a piecewise linear force-free field (e.g., Longcope and Malanushenko, 2008).  So, provided the initial state is a nonlinear force-free field, there will certainly exist a lower energy state with the latter connections that preserves magnetic helicity.

The second minor point is that, since the initial states are untwisted and the final states are twisted, it may be thought at first that the final state must have a higher energy than the initial state. But that conclusion is valid only for tubes of a given constant cross-sectional radius, and will not necessarily hold for tubes that are allowed to expand as they arch up from their footpoints.

\subsubsection{General Aspects}
\label{sec3.3.1}
The free magnetic energy (i.e.,\ the energy above potential) can be written as a sum over terms involving the total currents, $I_i$, flowing along their paths

\cite[\S 5.17]{jackson99},

\begin{eqnarray}
  \Delta W ~=~ W ~-~ W_0 ~=~ 
 \frac{1}{2\mu} \int\Bigl| {\bf B} - {\bf B}_0\Bigr|^2\, d^3 x ~=~ \half \sum_i L_i I_i^2 + \half \sum_i\sum_{j\ne i} M_{ij}I_iI_j ~~,
 \nonumber
\end{eqnarray}
where the coefficients $L_i$ and $M_{ij}$ depend on the current paths and the distribution of current therein.  Non-negativity 
of this energy requires that $L_i\ge 0$, for each current system $i$, and that $|M_{ij}|\le \sqrt{ L_i L_j }$ for all pairs $i$ and $j$.

In the initial configuration, shown schematically in Fig.\ 3a, separating domains $F_A$ from $F_B$  is a single current sheet carrying current $I_{\rm cs}$.  There is, at this time,  no current flowing through the domains, which are assumed untwisted.  As reconnection proceeds, current does increase in domains $F_R$ and $F_U$, which were also current-free before reconnection.  Changes $\delta I_{\rm cs}$, $\delta I_R$, and $\delta I_U$, to those currents change the free energy by an amount which is, to leading order,
\begin{eqnarray}
  \delta(\Delta W) ~=~ \Bigl(L_{\rm cs}\,\delta I_{\rm cs} ~+~  M_{{\rm cs},U}\,\delta I_U
  ~+~  M_{{\rm cs},R}\, \delta I_R\,\Bigr)\,I_{\rm cs} ~~.
   \nonumber
\end{eqnarray}
Reconnection will decrease the current in the sheet, so $\delta I_{\rm cs}\, I_{\rm cs}<0$.  At the same time the new domains, $U$ and $R$, are created with twisted flux, and so they have current.  It is because the coefficients $M_{{\rm cs},R}$ and $M_{{\rm cs},U}$ are so much smaller than $L_{\rm cs}$ that the
process of reconnection at a current sheet is energetically favourable ($\delta(\Delta W)<0$).  This state of affairs obtains during the initial phase, where contributions from $L_{R}\,I_R\,\delta I_R$ and $L_{U}\,I_U\,\delta I_U$ are negligible.  After a significant period of reconnection, however, these contributions, both necessarily positive, will eventually balance the negative contribution from the diminishing current sheet.  Reconnection will cease at such a point, with some portion of flux unreconnected, and a current sheet remaining.  We have not been able to determine this point in our simple model, and so instead have considered the limiting case where reconnection proceeds to completion.  This therefore provides an upper bound on helicity transfer.

\subsubsection{Simple Model}
\label{sec3.3.2}

Consider the two flux tubes in Fig. \ref{fig3}a stretching from A$_+$ to A$_-$ and B$_+$ to B$_-$, say, and suppose the flux tubes expand into the corona from small sources at these footpoints. A cylindrical tube of uniform twist $\Phi$, length $l$, radius $a$ and central axial field $B_0$, will have an equilibrium field of the Gold-Hoyle form \cite{priest14a}.  The field components in cylindrical polar coordinates ($R$,$\phi$,$z$) are
\begin{equation}
B_z=\frac{B_0}{1+q^2R^2}, \ \ \ \ \ B_\phi=\frac{B_0qR}{1+q^2R^2},
\label{eq19}
\end{equation}
where $q$ is a constant such that the twist is $\Phi=lq$. The net magnetic flux in such a tube is
\begin{equation}
F_0=\frac{\pi B_0}{q^2}\log{(1+q^2a^2)},
\label{eq20}
\end{equation}
while the magnetic energy is
\begin{equation}
W=\frac{l B_0 F_0}{2\mu}.
\label{eq21}
\end{equation}

We shall suppose for simplicity that our flux tubes expand rapidly up to form relatively uniform tubes in the corona, for which the flux ($F_0$) and twist ($\Phi$) are given. If these are held constant, then, as the radius $a$ increases and the tube expands more into the corona, so the central axial field $B_0$ decreases and the energy $W$ decreases. 

For the initial state in Fig. \ref{fig3}a, the length of the coronal part of each uniform untwisted flux tube is roughly $l_T=\sqrt(s^2+w^2)$ and the central axial field is $B_{0T}=F_0/(\pi a_T^2)$, where the geometry of the situation with its similar triangles implies that the radius of the tube (namely, half the perpendicular distance between the axes of the two tubes) is $a_T=\half w L/\sqrt(s^2+w^2)$. Thus, the central axial field can be written
\begin{equation}
B_{0T}=\frac{4 F_0 (w^2+s^2)}{\pi w^2 L^2}.
\label{eq22}
\end{equation}

For the final state in Fig. \ref{fig3}b, we suppose the tubes are twisted, with the value of $q$ and therefore the twist $\Phi=lq$ given by helicity conservation, so that in this case the axis field is given from Eq.(\ref{eq20}) by the form
\begin{equation}
B_0=\frac{F_0 q^2}{\pi \log(1+q^2a^2)}.
\label{eq23}
\end{equation}
For the overlying flux rope the length is given by roughly $ l_R =  \sqrt \{(L+s)^2+w^2\} $, while the geometry of the configuration implies that the radius of the tube (namely, roughly  the perpendicular distance between the tube axis and the footpoint of the underlying tube) has increased to $a_R=L w / \sqrt \{(L+s)^2+w^2\}$. The central axial field then becomes
\begin{equation}
B_{0R}=\frac{F_0 q_R^2}{\pi \log\{1+q_R^2L^2w^2/[(L+s)^2+w^2]\}}.
\label{eq24}
\end{equation}
The underlying flux loop has a rough length of $ l_U =  \sqrt \{(L-s)^2+w^2\} $ and a radius (namely, roughly half the perpendicular distance between the tube axis and the footpoint of the overlying tube) that has increased to $a_U= L w / \sqrt \{(L-s)^2+w^2\}$, 
so that its central axial field becomes
\begin{equation}
B_{0U}=\frac{F_0 q_U^2}{\pi \log\{1+q_U^2L^2w^2/[(L-s)^2+w^2]\}}.
\label{eq25}
\end{equation}

\begin{figure}[h]
{\centering
\includegraphics[width=12cm]{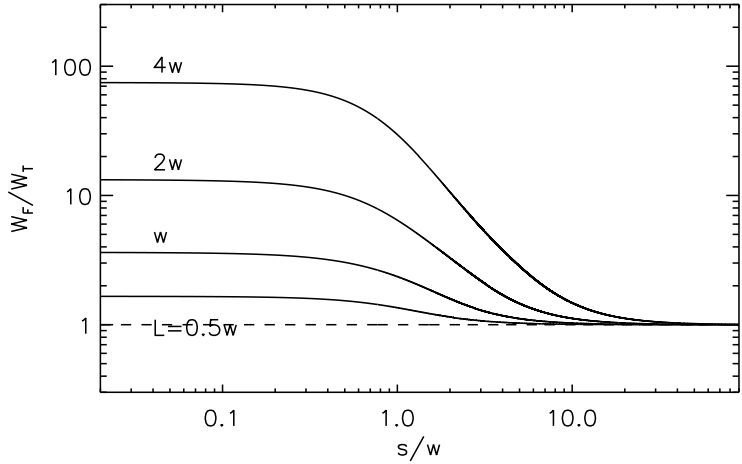}
\caption{The ratio ($W_F/W_T$) of the final to the initial energy for the loops shown in Fig. \ref{fig3} as a function of $s/w$ for varying $L/w$, where $s$ is the shear, while $w$ and $L$ are the separations shown in Fig. \ref{fig3} between sources of opposite and like polarity, respectively.}
\label{fig8}}
\end{figure}
Thus, using Eq. (\ref{eq21}), the initial energy becomes
\begin{equation}
W_T=\frac{4F_0^2(w^2+s^2)^{3/2}}{\pi \mu w^2 L^2},
\label{eq26}
\end{equation}
while the final energy is the sum of the energies of the overlying flux rope and underlying loops, namely,
\begin{eqnarray}
W_F\equiv W_R+W_U=\frac{\{(L+s)^2+w^2\}^{1/2}F_0^2q_R^2}{2\pi \mu\log\{1+q_R^2L^2w^2/[(s+L)^2+w^2]\}}+\nonumber \\
\frac{\{(L-s)^2+w^2\}^{1/2}F_0^2q_U^2}{2\pi \mu\log\{1+q_U^2L^2w^2/[(s-L)^2+w^2]\}}
\label{eq27}
\end{eqnarray}
The ratio ($W_F/W_T$) of final to initial energy is plotted as a function of $s/w$ for varying $L/w$ in Fig. \ref{fig8}, which shows that, when $s/w$ is large enough, $W_F/W_T<1$, so that the initial magnetic energy does indeed exceed the final energy, as required for an accessible state. In particular case when $sL\ll w$ and $qL\ll 1$, the initial and final energies are
\begin{equation}
W_T= \frac{4F_0^2w}{\pi \mu L^2}  \ \ \ \ \ {\rm and} \ \ \ \ \ W_F= \frac{F_0^2w}{\pi \mu L^2},
\label{eq28}
\end{equation}
so that $W_F/W_T=1/4$.  In future we plan to improve upon this energy calculation by considering partial reconnection and also taking account of the current sheet that is likely to be present, but both of those are outside the scope of the present paper.

\section{Modelling the Eruption of a Magnetic Arcade Containing a Flux Rope} 
\label{sec4}

Consider next a twisted magnetic flux rope  of initial twist $\Phi_{Ri}$ and flux $F_I$ situated under a coronal arcade of flux $2F_a$, and suppose that it reconnects with the arcade to produce an erupting flux rope whose core is the original flux rope, but which is now enveloped by a sheath of extra flux $F_a$ and twist $\Phi_{R}$, as sketched in Fig.\ref{fig2}. We here develop a simple model to determine $\Phi_{R}$ in terms of the geometry of the initial state. 

\begin{figure}[h]
{\centering
 \includegraphics[width=12cm]{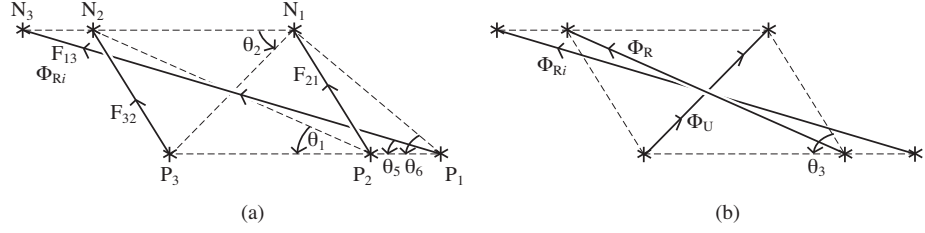}
\caption{A simple model for the eruption and reconnection of a flux rope (of flux $F_{12}$ and mean twist ${\bar \Phi}_{Ri}$) through an overlying coronal arcade (of flux $F_{21}$), showing (a) the pre-reconnection and (b) the post-reconnection configuration.}
\label{fig9}}
\end{figure}

Suppose the magnetic flux ($F_{13}=F_I$) of the flux rope  stretches from a positive source P$_1$ to a negative source N$_3$, while the overlying  arcade has two parts of flux $F_{21}=F_a$ and $F_{32}=F_a$ linking  positive sources P$_2$ and P$_3$ to negative sources N$_1$ and N$_2$, respectively (Fig.\ref{fig9}a). The arcade sources lie at the vertices of a parallelogram, with angles $\theta_1 ,\ \theta_2,\ \theta_3$ (equal to the angles N$_2$P$_2$P$_3$, N$_2$N$_1$P$_3$, N$_1$P$_2$P$_3$, respectively), while the flux rope source P$_1$ subtends angles $\theta_5$ and $\theta_6$, as indicated in Fig.\ref{fig9}.  After reconnection, the original flux rope still links P$_1$ and N$_3$, while the initial arcade has now reconnected, so that P$_2$ now joins to N$_2$ to give flux that winds round (with twist $\Phi_R$) and enhances the original flux rope, while P$_3$ now joins to N$_1$ to give an underlying set of loops with self-helicity $\Phi_UF_{31}^2/(2\pi)$, say (Fig.\ref{fig9}b).

The initial self-helicity of the flux rope is ${\Phi}_{Ri}F_{13}^2/(2\pi)$ and is preserved as the self-helicity of the core of the erupting flux rope after reconnection. The final self-helicity of the new part P$_2$N$_2$ of the erupting flux rope is ${\Phi}_{R}F_{22}^2/(2\pi)$, while the final self-helicity of the underlying loops P$_3$N$_1$ is ${\Phi}_{U}F_{31}^2/(2\pi)$. 

The initial mutual helicity has three parts, namely: $[1-\half(\theta_6+\theta_7)/\pi]F_{13}F_{21}$ due to P$_1$N$_3$ lying under P$_2$N$_1$, where $\theta_6$ is the angle N$_1$P$_1$P$_2$ and $\theta_7$ is the angle P$_2$N$_3$N$_1$;  $[1-\half(\theta_5+\theta_8)/\pi]F_{13}F_{32}$ due to P$_1$N$_3$ lying under P$_3$N$_2$, where $\theta_5$ is the angle N$_2$P$_1$P$_3$ and $\theta_8$ is the angle P$_3$N$_3$N$_2$;  and $[(\theta_2-\theta_1)/\pi] F_{21}F_{32}$ due to P$_3$N$_2$ lying alongside P$_2$N$_1$, where $\theta_1$ is the angle P$_3$P$_2$N$_2$ and $\theta_2$ is the angle P$_3$N$_1$N$_2$.
 
 After reconnection, the mutual helicity decreases to the sum of three parts, namely: $-[\half(\theta_6+\theta_7)/\pi]F_{13}F_{31}$ due to the initial flux rope P$_1$N$_3$ now lies over the underlying arcade of loops P$_3$N$_1$;  $-[\half(\theta_5+\theta_8)/\pi]F_{13}F_{22}$ due to the initial flux rope P$_1$N$_3$ lying under  the new part P$_2$N$_2$ of the erupting flux rope; and $-(\theta_3/\pi)F_{31}F_{22}$ due to the underlying arcade of loops P$_3$N$_1$ lying under  the new part P$_2$N$_2$ of the erupting flux rope.
 
We first assume {\it magnetic helicity equipartition},  so that the self-helicities  added to the underlying arcade and flux  rope are equal and the released mutual helicity is shared equally between the two flux tubes, i.e.,
\begin{eqnarray}
\frac{{\Phi}_{U}}{2\pi}F_{31}^2 = \frac{{\Phi}_{R}}{2\pi}F_{22}^2 .
\label{eq29}
\end{eqnarray}
 
We also assume {\it magnetic helicity conservation} so that the initial and final helicities (self plus mutual) are the same, namely
\begin{eqnarray}
\frac{{\Phi}_{Ri}}{2\pi}F_{13}^2 + \left(1-\frac{\theta_6+\theta_7}{2\pi}\right)F_{13}F_{21}+\left(1-\frac{\theta_5+\theta_8}{2\pi}\right)F_{13}F_{32}+\frac{\theta_2-\theta_1}{\pi}F_{21}F_{32}\nonumber \\ 
=\frac{{\Phi}_{Ri}}{2\pi}F_{13}^2 + \frac{{\Phi}_{R}}{2\pi}F_{22}^2+\frac{{\Phi}_{U}}{2\pi}F_{31}^2
-\frac{\theta_6+\theta_7}{2\pi}F_{13}F_{21}-\frac{\theta_5+\theta_8}{2\pi}F_{13}F_{32}-\frac{2\theta_3}{\pi}F_{21}F_{32}.
\nonumber
\end{eqnarray}
After substituting for $\Phi_U$ from Eq.(\ref{eq29}) and putting $F_{32}=F_{21}=F_{31}=F_{22}=F_a$, $F_{13}=F_I$, this reduces to
\begin{eqnarray}
{\Phi}_{R}=2\pi{\frac{F_I}{F_a}}+ \theta_2-\theta_1+2\theta_3.
\label{eq30}
\end{eqnarray}
Thus, by comparison with the case where there is no initial flux rope, the reconnection enhances the new flux rope twist by $2\pi F_I/F_a$.

\section{Discussion} 
\label{sec5}

We have set up a  simple model for estimating the twist in erupting prominences, in association with eruptive two-ribbon flares and/or with coronal mass ejections.  It is based on three simple assumptions, namely, conservation of magnetic flux, conservation of magnetic helicity and equipartition of magnetic helicity.  While the first and second are well established, the third is more of a reasonable conjecture.  In future, it would be interesting to test the model and the conjecture with both observations and computational experiments.

During the main phase of a flare, the shear of the flare loops is observed to decrease in time, so that they become oriented more perpendicular to the polarity inversion line. This is a natural consequence of our model, where flux is added to the flux rope first from the innermost parts of the arcade (i.e., closest to the polarity inversion line), so that, as can be seen in Figs.\ref{fig1}b, \ref{fig2}b and \ref{fig9}b, the final shear of the arcade is smaller than the initial shear.  In other words, the change in shear is a consequence of the geometry of the three-dimensional reconnection process.

The cause of the eruption is a separate topic that has been discussed extensively elsewhere (e.g., \opencite{priest14a}), and includes either nonequilibrium, kink instability, torus instability or breakout.  One puzzle is what happens with confined flares, where the flare loops and H$\alpha$ ribbons form but there is no eruption.  A distinct possibility is that the overlying magnetic field and flux are too strong to allow the eruption, but this needs to be tested by comparing nonlinear force-free extrapolations with observations (e.g., \opencite{wiegelmann08b,mackay11,mackay12b}). Another puzzle is the cause of preflare heating. One possibility is the slow initiation of reconnection before a fast phase (Yuhong Fan, private communication), but another is that the flux rope goes unstable to kink instability which spreads the heating nonlinearly throughout the flux rope in a multitude of secondary current sheets \cite{hood09a}; if the surrounding field is stable enough, such an instability can possibly occur without an accompanying eruption.

The present simple model can be developed in several ways, which we hope to pursue in future.  One is to conduct computational experiments, in which the energies before and after reconnection will be calculated in order to check which states are energetically accessible.  A  second way is to extend the model to more realistic initial configurations with more elements, in which the reconfiguration is by quasi-separator or separator reconnection \cite{priest96a,longcope96,longcope05a,longcope2008b,parnell10b} and the internal structure of the flux rope and arcade are taken into account.  In particular, the distribution of magnetic flux within the arcade will be included, both normal to and parallel to the polarity inversion line.

\appendix
\section
{Details of Magnetic Helicity Conservation}

In Eq.(\ref{eq9}),  the mutual helicities may be written using Eq.(\ref{eq2}) as
\BA
H_{AB}&=& \frac{(\theta_2^{AB}-\theta_1^{AB})}{2\pi}(F_a-F)^2,\nonumber
\\
H_{AR}&=& \frac{(\theta_2^{AR}-\theta_1^{AR})}{2\pi}(F_a- F)F,\nonumber
\\
H_{AU}&=& \frac{(\theta_2^{AU}-\theta_1^{AU})}{2\pi}(F_a-F)F,\label{eq31}
\\
H_{BR}&=& \frac{(\theta_2^{BR}-\theta_1^{BR})}{2\pi}(F_a- F)F,\nonumber
\\
H_{BU}&=& \frac{(\theta_2^{BU}-\theta_1^{BU})}{2\pi}(F_a- F)F,\nonumber
\\
H_{RU}&=& \frac{-\theta_3}{\pi}F^2,\nonumber
\EA
where the angles may be written in terms of the angles between various lines joining the footpoints A$_+$, R$_+$, U$_+$, B$_+$,  A$_-$, R$_-$, U$_-$ and B$_-$ in Fig.\ref{fig5}d, namely, 
$\theta_1^{AB}=\angle B_+A_+B_-$, $\theta_2^{AB}=\angle B_+A_-B_-$, 
$\theta_1^{AR}=\angle B_+A_+R_-$, $\theta_2^{AR}=\angle R_+A_-B_-$, 
$\theta_1^{AU}=\angle B_+A_+U_-$, $\theta_2^{AU}=\angle U_+A_-B_-$, 
$\theta_1^{BR}=\angle B_+R_+B_-=\theta_1^{AR}$, $\theta_2^{BR}=\angle B_+R_-B_-=\theta_2^{AR}$, 
$\theta_1^{BU}=\angle B_+U_+B_-=\theta_1^{AU}$, $\theta_2^{BU}=\angle B_+U_-B_-=\theta_2^{AU}$  
and $\theta_3=\angle B_+A_+A_-$.

In a similar way to Eq.(\ref{eq5}) for $\theta_1$ and $\theta_2$, these angles may also be written in terms of $\bar{F}$ and the geometrical parameters $\bar{L}$ and $\bar{s}$ as
\begin{eqnarray}
\tan \theta_1^{AB}&=&\frac{w}{L+L\bar{F}+s}=\frac{1}{\bar{L}+\bar{L}\bar{F}+\bar{s}}, \nonumber \\
\tan \theta_2^{AB}&=&\frac{w}{L+L\bar{F}-s}=\frac{1}{\bar{L}+\bar{L}\bar{F}-\bar{s}}, \nonumber \\
\tan \theta_1^{AR}&=&\frac{w}{s+(\half+\bar{F})L}=\frac{1}{\bar{s}+(\half+\bar{F})\bar{L}}, \nonumber \\
 \tan \theta_2^{AR}&=&\frac{w}{\half L-s}=\frac{1}{\half\bar{L}-\bar{s}}, \label{eq32}\\
\tan \theta_1^{AU}&=&\frac{w}{\half L+s}=\frac{1}{\half\bar{L}+\bar{s}}, \nonumber \\
 \tan \theta_2^{AU}&=&\frac{w}{-s+(\half+\bar{F})L}=\frac{1}{-\bar{s}+(\half+\bar{F})\bar{L}},  \nonumber \\
 \tan \theta_3&=&\frac{w}{s}=\frac{1}{\bar{s}},\nonumber
\end{eqnarray}

In Eq.(\ref{eq10}) for the initial helicity, the mutual helicities are
\BA
H^{i}_{AB}&=& \frac{(\theta_2^{AB}-\theta_1^{AB})}{2\pi}(F_a-F)^2,\nonumber
\\
H^i_{AR}&=& \frac{(\theta_2^{AR}-\theta_1^{AU})}{2\pi}(F_a- F)F,\nonumber
\\
H^i_{AU}&=& \frac{(\theta_2^{AU}-\theta_1^{AR})}{2\pi}(F_a-F)F,\nonumber
\\
H^i_{BR}&=& \frac{(\theta_2^{BU}-\theta_1^{BR})}{2\pi}(F_a- F)F,\label{eq33}
\\
H^i_{BU}&=& \frac{(\theta_2^{BR}-\theta_1^{BU})}{2\pi}(F_a- F)F,\nonumber
\\
H^i_{RU}&=& \frac{(\theta_2^{RU}-\theta_1^{RU})}{2\pi}F^2.\nonumber
\EA

Thus, we note that, although all of the mutual helicities change during the course of the reconnection, the sums
$H_{AR}+H_{AU}$ and $H_{BR}+H_{BU}$ remain the same.

By substituting these into the expressions for initial and final helicity (Eqs.\ref{eq9}, \ref{eq10}), helicity conservation gives the mean flux rope twist in Eq.(\ref{eq12}), as required.

\acknowledgements
We are grateful to Mitch Berger, Pascal D{\' e}moulin, Alan Hood and Clare Parnell for helpful comments and suggestions and to the UK STFC, High Altitude Observatory and Montana State University for financial support.

\bibliographystyle{sp}

\bibliography{Helicity}

\end{article}
\end{document}